\documentclass[aps,prl,twocolumn,superscriptaddress,floatfix]{revtex4-1}
\usepackage{graphicx}
\usepackage{blindtext}
\usepackage{lipsum}
\usepackage{color}
\usepackage[dvipsnames]{xcolor}
\usepackage{amsmath}
\usepackage{amssymb}
\usepackage{wasysym}
\usepackage{epstopdf}
\usepackage{fdsymbol}
\usepackage{bm}
\usepackage{pdfpages}
\usepackage{appendix}
\usepackage{pgffor}

\newif\ifarXiv
\arXivtrue 

\makeatletter
\AtBeginDocument{\let\LS@rot\@undefined}
\makeatother

\begin{document}


\title{Self-assembly of Freely-rotating Polydisperse Cuboids:\\ Unveiling the Boundaries of the Biaxial Nematic Phase}


\author{Effran Mirzad Rafael}
\author{Daniel Corbett}
\affiliation{Department of Chemical Engineering and Analytical Science, The University of Manchester, Manchester, M13 9PL, United Kingdom}
\author{Alejandro Cuetos}
\affiliation{Department of Physical, Chemical and Natural Systems, Pablo de Olavide University, 41013 Sevilla, Spain}
\author{Alessandro Patti}
\email{alessandro.patti@manchester.ac.uk}
\affiliation{Department of Chemical Engineering and Analytical Science, The University of Manchester, Manchester, M13 9PL, United Kingdom}



\begin{abstract}
Colloidal cuboids have the potential to self-assemble into biaxial liquid crystal phases, which exhibit two independent optical axes. Over the last few decades, several theoretical works predicted the existence of a wide region of the phase diagram where the biaxial nematic phase would be stable, but imposed rather strong constraints on the particle rotational degrees of freedom. In this work, we employ molecular simulation to investigate the impact of size dispersity on the phase behaviour of freely-rotating hard cuboids, here modelled as  self-dual-shaped nanoboards. This peculiar anisotropy, exactly in between oblate and prolate geometry, has been proposed as the most appropriate to promote phase biaxiality. We observe that size dispersity radically changes the phase behaviour of monodisperse systems and leads to the formation of the elusive biaxial nematic phase, being found in an large region of the packing fraction \textit{vs} polydispersity phase diagram. Although our results confirm the tendencies reported in past experimental observations on colloidal dispersions of slightly prolate goethite particles, they cannot reproduce the direct isotropic-to-biaxial nematic phase transition observed in these experiments 
\end{abstract}


\maketitle


The first known theory on entropy-driven phase transitions was proposed by Onsager in his seminal work dating back to 1940s \cite{onsager1949effects}. Onsager demonstrated that systems of infinitely long hard rods exhibit isotropic-to-nematic phase transition as a result of mere volume effects. Later on, experiments and computer simulations showed that entropy-driven phase transitions can also lead to the formation of positionally ordered liquid crystals (LCs), such as smectic and columnar phases \cite{frenkel1988thermodynamic, van2000liquid}. The equilibrium structures stemming from the self-assembly of colloidal particles are especially determined by the architecture of their building blocks. In particular, biaxial particles, such as bent-core and cuboidal particles, have been reported to form biaxial nematic ($\rm N_{B}$) LCs \cite{acharya2004biaxial, madsen2004thermotropic, van2009experimental}. First theorised by Freiser in 1970 \cite{freiser1970ordered}, the $\rm N_{B}$ phase has attracted widespread attention for being a promising candidate to be engineered into next generation liquid crystal displays. In contrast to uniaxial nematics ($\rm N_{U}$), where long-range orientational order exists only along one direction, the $\rm N_{B}$ phase possesses three orthogonal directors and hence two distinct optical axes that can pave the path to high-performance displays \cite{lee2007dynamics, berardi2008field, ricci2015field}. Despite having been extensively studied over the past 50 years, the stability of the $\rm N_{B}$ phase still remains an open question, predominantly, but not only, at the molecular scale. Answering this question is hampered by the fact that the $\rm N_{B}$ phase tends to be metastable with respect to other morphologies, such as the $\rm N_{U}$ and $\rm Sm$ phases \cite{taylor1991nematic}. 

About a decade ago, Vroege and coworkers reported on the first experimental evidence of the $\rm N_{B}$ phase in a system of colloidal goethite (roughly board-like) particles \cite{van2009experimental}. The stability of this $\rm N_{B}$ phase was ascribed to the particles' quasi self-dual shape, a geometry in between oblate and prolate, and to their significant size dispersity, which hinders the formation of the $\rm Sm$ phase. This key work has reignited recent interest, sparking numerous theoretical, experimental and computer simulation studies on hard board-like particles (HBPs)  \cite{martinez2011biaxial, belli2011polydispersity, belli2012depletion, peroukidis2013phase, peroukidis2013supramolecular, peroukidis2014biaxial, op2014tuning, martinez2014phase, mederos2014hard, gonzalez2015effect, cuetos2017phase, patti2018monte, dussi2018hard, yang2018synthesis, cuetos2019biaxial, skutnik2020biaxial} and other biaxial geometries \cite{tasios2017simulation, querciagrossa2017, orlandi2018, dussi2018hard, drwenski2018effect, chiappini2019biaxial}. Theoretical and computational studies on monodisperse systems have suggested that self-dual-shaped particles exhibit a higher tendency to form biaxial nematics. However, these studies applied rather strong approximations, limiting the particles orientation to six orthogonal directions \cite{taylor1991nematic, martinez2011biaxial, belli2011polydispersity}, freezing the rotation of the particle long axes \cite{vanakaras2003theory}, or neglecting the occurrence of positionally ordered LC phases \cite{straley1974ordered, mulder1989isotropic, allen1990computer, camp1997phase}. Our recent theoretical work and computer simulations of freely-rotating HBPs suggested that these approximations might artificially magnify the stability of the $\rm N_{B}$ phase \cite{cuetos2017phase, patti2018monte}. We note that stable $\rm N_{B}$ phases have been found in systems of cuboids with rounded corners (spheroplatelets) with length-to-thickness ratio $L^* \equiv L/T>9$ \cite{peroukidis2013supramolecular, peroukidis2014biaxial} and in systems of especially elongated HBPs ($L^* \geq 23$) \cite{dussi2018hard}. Nevertheless, experiments on highly uniform and monodisperse colloidal cuboids with $15 \le L^* \le 180$ did not report the formation of the $\rm N_{B}$ phase, which, for self-dual-shaped cuboids, was found to be pre-empted by the biaxial smectic ($\rm Sm_{B}$) phase \cite{yang2018synthesis}. This lack of agreement between experiments, theory and simulations keeps the discussion on the ability of HBPs to form the $\rm N_{B}$ phase still alive.  

Size dispersity has been identified as a key ingredient to destabilise the Sm phase and thus promote the formation of biaxial nematics. In particular, the effect of size dispersity in systems of HBPs was investigated by Onsager's theory within the restricted-orientation (Zwanzig) approximation \cite{belli2011polydispersity, op2014tuning}. This theory provided an elegant and solid explanation on the origin of the $\rm N_{B}$ stability experimentally observed in colloidal dispersions of goethite particles \cite{van2009experimental}. Nevertheless, its conclusions were strongly determined by the use of the Zwanzig model, which only allows six orthogonal particle orientations and cannot describe the phase behaviour of cuboids accurately, as recent simulations and theory have indicated \cite{cuetos2017phase, patti2018monte}. Consequently, fully unlocking the particle rotational degrees of freedom is of paramount importance to ascertain the impact of polydispersity on the phase behaviour of HBPs and accurately map the boundaries of the $\rm N_{B}$ phase. While it is extremely challenging to formulate a theory that simultaneously incorporates particle size dispersity and unrestricted orientations, molecular simulation can provide an insightful contribution to shed light on this combined effect. To this end, we have performed Monte Carlo (MC) simulations of freely-rotating HBPs with Gaussian size distribution peaked at $L^* = 12$. The particle thickness, $T$, is the system unit length and is the same for all HBPs, whereas $L^{*}$ changes with standard deviation $\sigma_L \left< L^* \right>$, where $0.05 \le \sigma_L \le 0.30$ measures the particle length dispersity. Finally, the width-to-thickness ratio, $W^{*} \equiv W/T = \sqrt{L^*}$, sets the self-dual shape for all particles. Our systems consist of $N_p = 2000$ to $3000$ HBPs that are initially arranged in cubic or rectangular boxes with periodic boundaries and are equilibrated in the isothermal-isobaric ensemble. Phase transitions have been assessed by expansion or compression of a perfect biaxial nematic phase at an extensive range of pressures. To ensure that the equilibrium configurations were independent of the initial configurations, expansion of $\rm Sm$ phases and compression of isotropic (I) phases have also been carried out. We have also simulated significantly larger systems, with $N_p=6000$, to discard the occurrence of finite-size effects, where these could especially influence the symmetry of the phases observed, that is at the $\rm I-N$ transition. Generally, up to $10^{7}$ MC cycles were needed to equilibrate the systems, with a cycle consisting of $N_p$ attempts of displacing and/or rotating randomly selected particles and one trial volume change. Because the force field employed here only consists of a hard-core potential, these moves were accepted if no overlaps were detected, according to the separating axes theorem \cite{gottschalk1996obbtree, john2008phase}. Systems were considered at equilibrium if packing fraction ($\eta$) and uniaxial ($S_{2}$) and biaxial ($B_{2}$) order parameters achieved steady values within reasonable statistical uncertainty. In particular, $\eta \equiv \sum_{i=1}^{N_p} v_{i} /V$, where $V$ is the box volume and $v_{i}$ the volume of a generic particle $i$. The calculation of $S_{2}$ and $B_{2}$ was done by diagonalisation of the traceless second-rank symmetric tensor $\textbf{Q}^{\lambda \lambda}= \langle \sum_{i=1}^{N_p} \big(3\hat{\lambda}_{i} \cdot \hat{\lambda}_{i} - \textbf{I}\big) \rangle/2N_p$, where $\bf I$ is the second-rank unit tensor, $\hat{\lambda}_{i}=\hat{x}, \hat{y}$ and $\hat{z}$ is the particle unit orientation vector and angular brackets denotes ensemble average. Isotropic, perfect uniaxial and perfect biaxial phases are observed at $(S_2, B_2)=(0,0)$, $(1,0)$ and $(1,1)$, respectively. Finally, to assess the long-range ordering of the phases at equilibrium, we have analysed the spatial correlations along the relevant phase directors and perpendicularly to them by computing the longitudinal, $g_{\parallel}(r_{\parallel})$, and transverse, $g_{\perp}(r_{\perp})$, pair distribution functions, where $r_{\parallel}$ and $r_{\perp}$ are the corresponding projections of the inter-particle distance. The interested reader is referred to the ESI$\dag$ for further details on the calculation of order parameters and pair correlation functions.

\begin{figure}[ht!]
    \centering
    \includegraphics[width=1.02\linewidth, height=0.282\textheight]{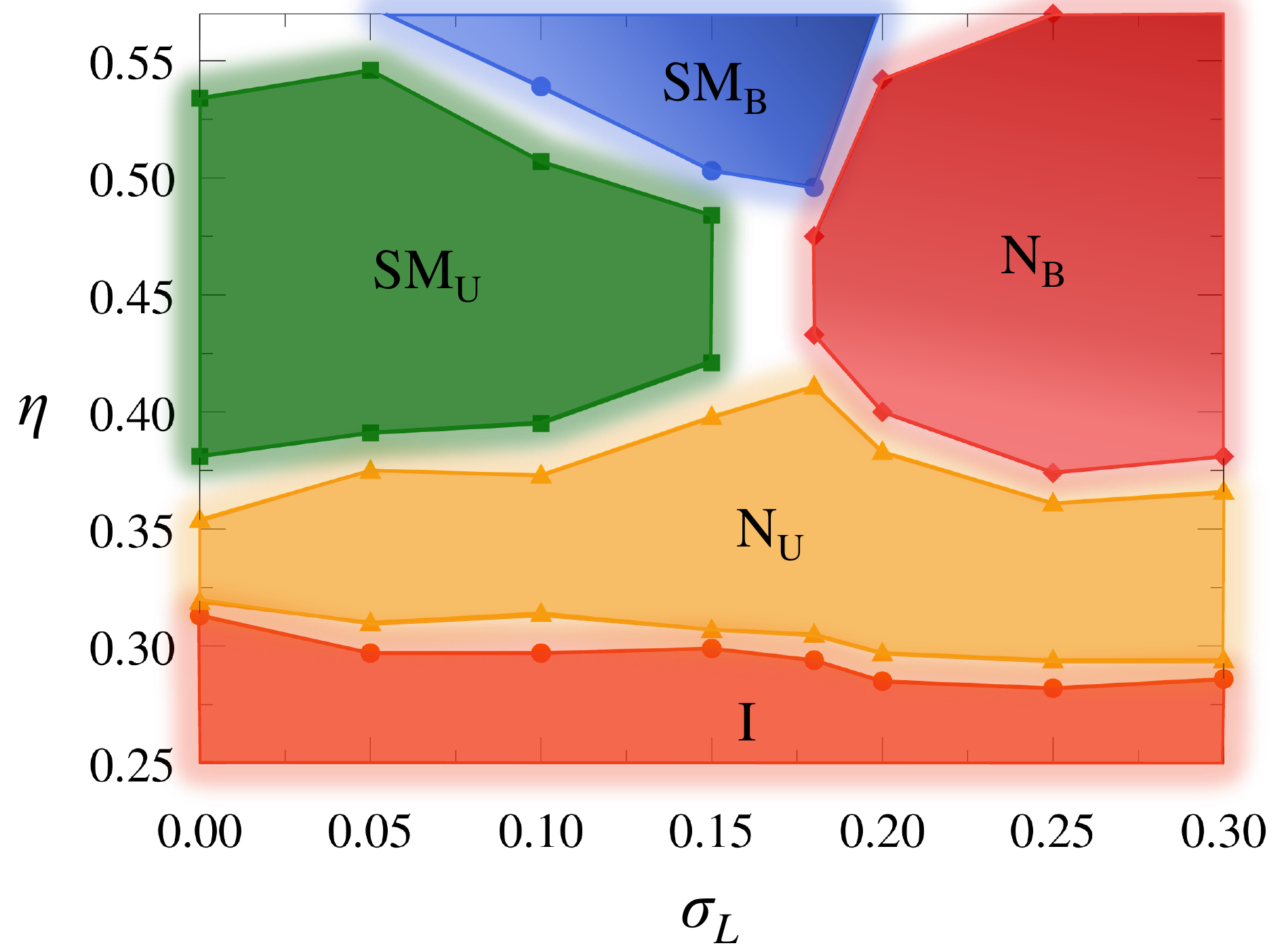}
    \caption{Phase diagram of polydisperse and freely-rotating HBPs with $\langle L^{*} \rangle = 12$, $\langle W^{*} \rangle = \sqrt{12}\approx 3.46$ and length polydispersity $0 \leq \sigma_{L} \leq 0.30$. The following phases are identified: I (\textcolor{RedOrange}{$\CIRCLE$}), N$_\text{U}$ (\textcolor{YellowOrange}{$\blacktriangle$}), N$_\text{B}$ (\textcolor{red}{$\medblackdiamond$}), Sm$_\text{U}$ (\textcolor{ForestGreen}{$\blacksquare$}), and Sm$_\text{B}$ (\textcolor{RoyalBlue}{$\CIRCLE$}).}
    \label{fig:phasediagram}
\end{figure}

\begin{figure*}[ht!]
\centering
    \includegraphics[width=\textwidth]{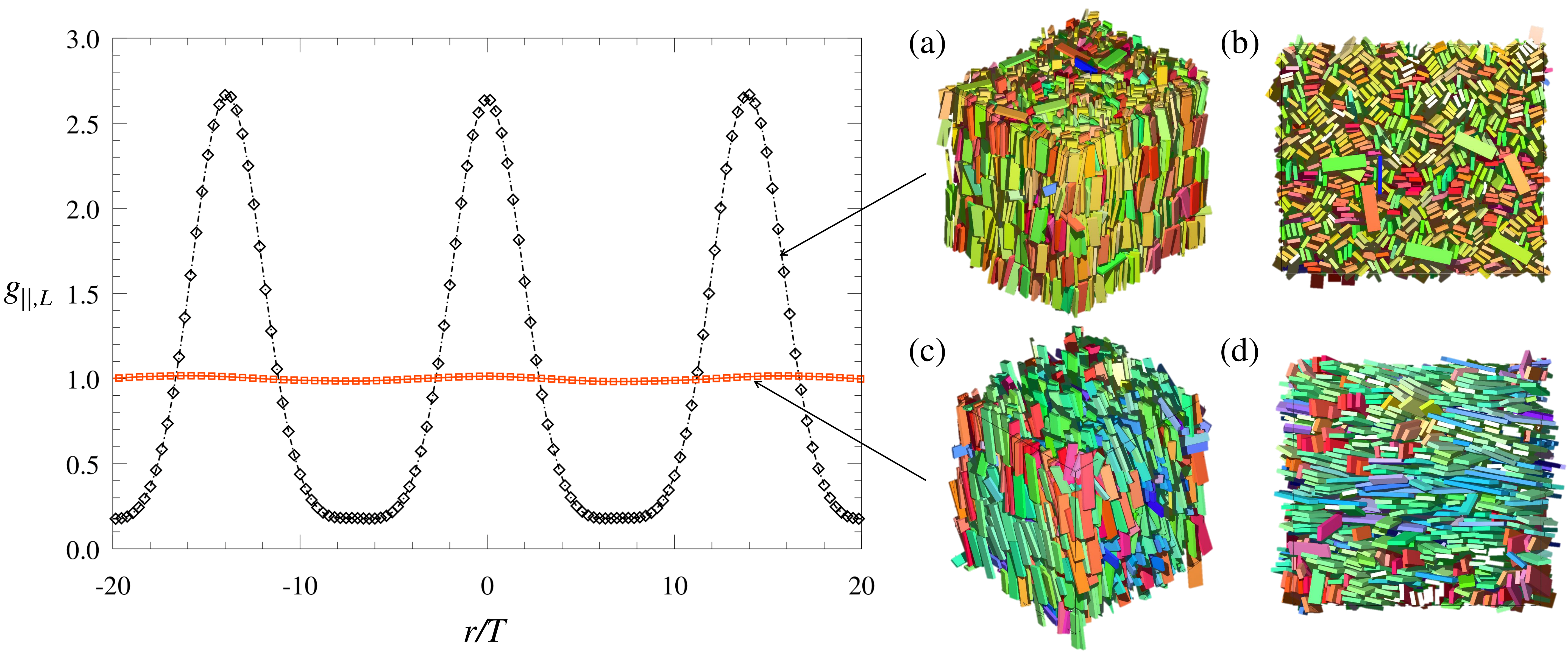}
   \caption{Parallel pair distribution functions along the nematic director of ($a,b$) a uniaxial smectic phase of prolate symmetry at $\sigma_{L} = 0.05$ and $\eta = 0.415$ (\textcolor{Black}{$\Diamond$}), and ($c,d$) a biaxial nematic phase at $\sigma_{L} = 0.25$ and $\eta = 0.423$ (\textcolor{RedOrange}{$\square$}). Different colours indicate different particle orientations.}
    \label{fig:gpar}
\end{figure*}


Bearing in mind these introductory considerations, we now report on the LC phases that polydisperse HBPs are able to form at equilibrium, with specific interest in the critical polydispersity that stabilises the $\rm N_{B}$ phase and the extent of this stability. For consistency with our former work \cite{cuetos2019biaxial}, we defined the axial symmetry of N and Sm phases according to the magnitude of the order parameters (see ESI$\dag$ for additional details). The $\sigma_{L} - \eta$ phase diagram of polydisperse and self-dual-shaped HBPs is shown in Fig.\ \ref{fig:phasediagram}. 
All points at $\sigma_L=0$ refer to monodisperse systems, obtained from previous simulation results \cite{cuetos2017phase} and included here as a reference. The diagram reveals the existence of uniaxial and biaxial LC phases at $\eta \gtrsim 0.30$. At lower density and regardless the size dispersity, we only detect I phases, in good agreement with the Zwanzig-based Onsager theory by Belli \textit{et al}, who predicted stable I phases for $\eta \lesssim 0.28$ and $0 \le \sigma_L \le 0.4$ \cite{belli2011polydispersity}. We stress that these authors studied slightly prolate particles with $W^* \approx 1.017 \sqrt{L^*}$ or $\nu \equiv L/W - W/T = 0.1$. This slight deviation from the self-dual shape, for which $\nu=0$, might appear insignificant, but in fact it determines the oblate or prolate symmetry of the nematic phase, at least in monodisperse systems, as established by Mulder in the 1980s \cite{mulder1989isotropic}.

As far as the $\rm N_{U}$ phase is concerned, its stability region (yellow-shaded area in Fig.\ \ref{fig:phasediagram}) is not especially influenced by the particle polydispersity, being the lower and upper boundaries, approximately constrained between $\eta=0.30$ and 0.35, similar to those observed in monodisperse systems ($\sigma_{L} = 0$). A slight difference is detected at $0.15 \le \sigma_{L} \le 0.20$, where four different phases seem to converge and the region of stability of uniaxial nematics expands up to $\eta=0.41$ at $\sigma_{L} = 0.18$. In former simulation studies on monodisperse systems of self-dual-shaped HBPs with $L^{*} = 12$, the $\rm N_{U}$ stability was also observed to be relatively small \cite{dussi2018hard}, and even vanish in binary mixtures \cite{patti2018monte}. Within the $\rm N_{U}$ domain, we came across nematic LCs with oblate and prolate symmetry, respectively labelled as $\rm N^-_{U}$ and $\rm N^+_{U}$. The occurrence of the $\rm N^-_{U}$ phase is predominantly detected close to the I-N transition and also far from it for $\sigma_L \ge 0.25$. By contrast, the $\rm N^+_{U}$ phase is mostly observed at larger packing fractions for $0 \le \sigma_L \le 0.25$. Consequently, at a given polydispersity, increasing the system density can produce an inversion of the nematic phase symmetry. While prolate and oblate particles commonly tend to trigger, respectively, $\rm N^+_{U}$ and $\rm N^-_{U}$ phases \cite{alben1973phase, straley1974ordered, mulder1989isotropic, taylor1991nematic, camp1997phase, op2014tuning}, although a significant polydispersity can counteract this tendency \cite{belli2011polydispersity}, the self-dual shape is in principle not expected to exhibit a clear oblate or prolate nature, but rather an ambivalent one. By applying our Onsager-like theory \cite{cuetos2017phase}, we observed that the free-energy difference between $\rm N^+_{U}$ and $\rm N^-_{U}$ phases is very small close to the I-N transition and not much larger at increasing $\eta$ (see ESI$\dag$). Although this theory is strictly valid for monodisperse systems and should serve here as a mere qualitative guideline, it confirms the slightly larger stability of the $\rm N^-_{U}$ in the vicinity of the I-N transition, in very good agreement with the tendencies observed in our simulations. Very small free-energy differences between oblate and prolate symmetries had also been reported by Mart{\'\i}nez-Rat{\'o}n and co-workers, who applied  density-functional theory to study the phase behaviour of nearly self-dual-shaped monodisperse HBPs \cite{martinez2011biaxial}.

\begin{figure*}[ht!]
\centering
    \includegraphics[width=\textwidth]{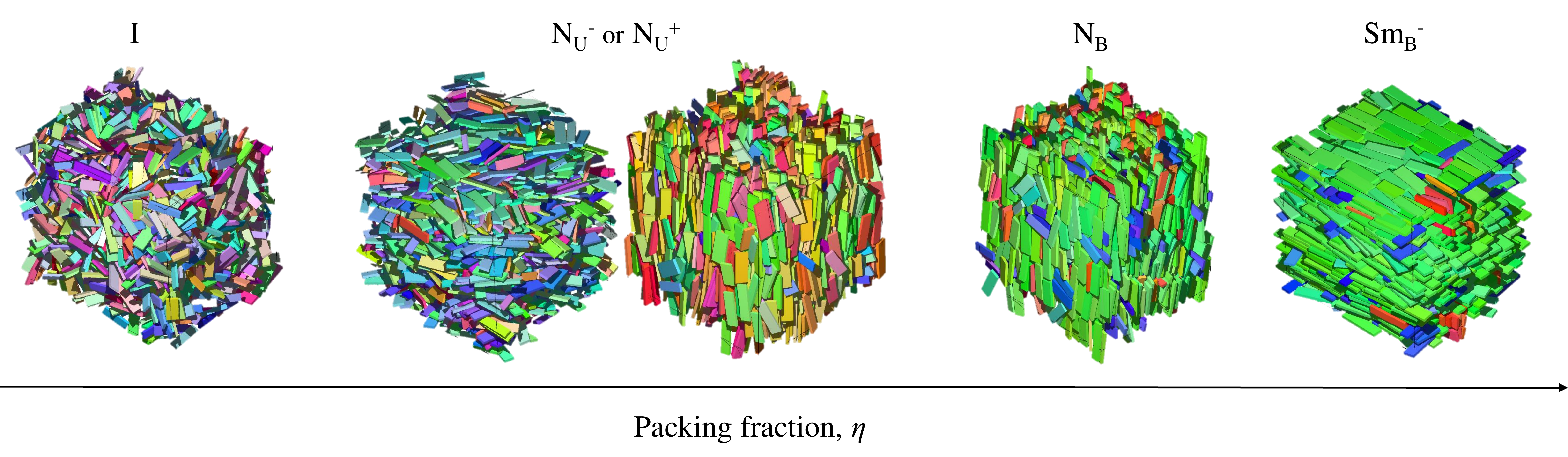}
   \caption{Equilibrium phases of HBPs at $\sigma_L = 0.18$ in an isotropic phase ($\eta = 0.266$), oblate nematic phase ($\eta = 0.314$), prolate nematic phase ($\eta = 0.341$), biaxial nematic phase ($\eta = 0.455$) and a biaxial smectic phase of oblate symmetry ($\eta = 0.573$). Different colours indicate different particle orientations.}
    \label{fig:sequence}
\end{figure*}

Uniaxial and biaxial smectic phases are obtained at larger packing fractions. To distinguish them from the nematic phases, we calculated the density distribution function along the nematic director, $g_{\parallel}(r_{\parallel})$, identified by the order parameters. Uniaxial smectic ($\rm Sm_{U}$) phases (green area in Fig.\ \ref{fig:phasediagram}) only exhibit a prolate symmetry ($\rm Sm^+_{U}$), in agreement with former simulations of monodisperse systems \cite{cuetos2017phase}. In particular, their layers, whose thickness is roughly $13T$ to $15T$, are perpendicular to the average direction of the particle length. The $g_{\parallel}(r_{\parallel})$, calculated along this direction, displays periodically peaked profiles of the type reported in Fig.\ \ref{fig:gpar} and no indication of structural order in the other directions. By contrast, biaxial smectics, found at $\eta > 0.50$ and $0.05 \le \sigma_{L} \le 0.20$ (blue area in Fig.\ \ref{fig:phasediagram}), can present a prolate ($\rm Sm^+_{B}$) or oblate ($\rm Sm^-_{B}$) symmetry. The former is characterised by layers piling along the particle length, while the latter by layers piling along the particle thickness. Upon increasing polydispersity, the $\rm Sm_{B}$ phases acquire a more and more defined structural identity, with a weak-to-strong biaxiality crossover at approximately $0.05 < \sigma_L < 0.10$ and a full biaxial character ($B_2>0.6$) at $\sigma_L \ge 0.18$. At this value of size dispersity, our simulations highlighted a particularly rich phase behaviour, unveiling an $\rm I \rightarrow N_{U} \rightarrow N_{B} \rightarrow Sm_{B}$ sequence of phases that is exemplary shown in Fig.\ \ref{fig:sequence}. In qualitative agreement with theory and experiments, we found the $\rm N_{B}$ phase to be stabilised by a substantial degree of particle size dispersity. In particular, our simulation results indicate $\sigma_{L} = 0.18$ as the critical polydispersity above which the $\rm N_{B}$ phase can form. Close to the $\rm N_{U}-N_{B}$ phase boundary, we find nematics with a relatively weak (but non-negligible) biaxiality and a residual oblate or prolate character. These phases, here referred to as weak biaxial nematics, are characterised by a biaxial order parameter in the range $0.2 \le B_2 \le 0.30$ and one predominant uniaxial order parameter granting them prolate ($\rm N_{B}^{+}$) or oblate ($\rm N_{B}^{-}$) symmetry.

In contrast with the experiments on goethite particles \cite{van2009experimental}, we do not observe a direct $\rm I-N_{B}$ phase transition here. This apparent lack of agreement deserves some comments. First of all, the cuboids studied in these experiments are not self-dual-shaped. Their shape parameter, $\nu = 0.1$, indicates a prolate geometry, which in monodisperse systems is expected to promote an $\rm I-N^+_{U}$, rather than $\rm I-N_{B}$, transition as predicted by theories spanning almost five decades \cite{alben1973phase, straley1974ordered, mulder1989isotropic, taylor1991nematic, camp1997phase, op2014tuning}. However, the goethite particles employed by Vroege and co-workers are not monodisperse, but exhibit a polydispersity between 20\% and 25\% in the three directions. Because polydispersity leads to fractionation \cite{van2008influence} and these authors studied the phase behaviour in capillaries, the longer particles tend to accumulate towards the bottom, where the $\rm N_{B}$ phase was found, \textit{de facto} increasing the shape parameter of this subset of particles to the effective value of $\nu = 0.6$ \cite{van2009experimental}. This particle geometry, evidently prolate, is very different from the self-dual shape applied here and a quantitative analogy is therefore not directly possible. Onsager theory within the Zwanzig approximation does not predict a direct $\rm I-N_{B}$ transition in systems of polydisperse HBPs with $\nu = 0.1$, but suggests the existence of the $\rm N_{B}$ phase in a wide region of the $\eta - \sigma_L$ phase diagram, including for $\sigma_L < 0.1$ \cite{belli2011polydispersity}. While it is known that restricting orientations can significantly enhance the stability of the $\rm N_{B}$ phase, both Belli's theoretical work and our simulations do not report a direct $\rm I-N_{B}$ phase transition, whose existence has never been unambiguously confirmed by off-lattice simulations spanning more than twenty years \cite{camp1997phase, berardi2008, preeti2011does, peroukidis2013supramolecular, querciagrossa2013, dussi2018hard, orlandi2018, querciagrossa2018}. Indeed, our recent MC simulations and generalised Onsager theory applied to freely-rotating monodisperse HBPs had even excluded the existence of the $\rm N_{B}$ phase, also at the self-dual shape\cite{cuetos2017phase}. To the best of our knowledge, there are no theoretical works on freely-rotating polydisperse HBPs that might help resolve this conundrum. While the phase diagram in Fig.\ \ref{fig:phasediagram} presents relevant discrepancies with that proposed by Belli, both works agree very well on the key role of polydispersity in the stabilisation of the $\rm N_{B}$ phase. This is especially  evident at $\sigma_{L} \ge 0.18$, where the stability region of the $\rm N_{B}$ phase widens, remarkably reducing that of $\rm Sm$ and $\rm N_{U}$ phases. This is not surprising as a large size dispersity is expected to hinder the formation of layered structures due to the absence of a well-defined structural periodicity in the longitudinal direction.

In summary, our MC simulations of freely-rotating HBPs have revealed a rich phase behaviour that is characterised by three key results: (\textit{i}) a significant degree of particle size dispersity is needed to stabilise the $\rm N_{B}$ phase; (\textit{ii}) self-dual-shaped HBPs do not exhibit direct $\rm I-N_{B}$ phase transition in the range of size dispersities studied here; (\textit{iii}) the ambivalent nature of the self-dual shape provides uniaxial nematics that, in a relatively wide region of the $\eta - \sigma_L$ phase diagram, might well be oblate or prolate. More specifically, a significant polydispersity ($\sigma_L \ge 0.18$) prevents the discretization of the space along the nematic director and thus enfeebles the stability of the Sm phase, practically enhancing that of the nematic phase. This result is in line with experiments and theory, but on a mere qualitative basis only. The lack of a direct $\rm I-N_{B}$ phase transition might appear in evident disagreement with previous experimental observations, which anyway employed prolate rather than self-dual-shaped particles, but agrees with a relevant number of simulation studies that could not confirm its occurrence in a reasonable range of particle anisotropies. As far as the ambivalence of the self-dual shape is concerned, it is interesting to observe an oblate-to-prolate symmetry inversion in the $\rm N_{U}$ domain upon increasing density. Our modified version of Onsager theory for monodisperse biaxial particles suggests that the free-energy difference between $\rm N^-_{U}$ and $\rm N^+_{U}$ phases would favour the former close to the $\rm I-N$ transition and the latter far from it.\\

EMR would like to thank the Malaysian Government Agency Majlis Amanah Rakyat for funding his PhD at the University of Manchester. AC acknowledges the Spanish Ministerio de Ciencia, Innovaci\'on y Universidades and FEDER for funding (project PGC2018-097151-B-I00) and C3UPO for the HPC facilities provided. AP acknowledges financial support from the Leverhulme Trust Research Project Grant RPG-2018-415. We would like to thank Andrew Masters and Matthew Dennison for sharing the code to calculate the virial coefficients. Finally, EMR, DC and AP acknowledge the assistance given by IT Services and the use of Computational Shared Facility at the University of Manchester.

\bibliographystyle{apsrev4-2}
\bibliography{preprint.bib}

\pagebreak



\section*{Supplementary material}

# Supplementary Information

## Self-assembly of Freely-rotating Polydisperse Cuboids: Unveiling the Boundaries of the Biaxial Nematic Phase

Effran Mirzad Rafael,[1] Daniel Corbett,[1] Alejandro Cuetos,[2] and Alessandro Patti[1]

[1]*Department of Chemical Engineering and Analytical Science,*
*The University of Manchester, Manchester, M13 9PL, United Kingdom*
[2]*Department of Physical, Chemical and Natural Systems,*
*Pablo de Olavide University, 41013 Sevilla, Spain*

## I. ORDER PARAMETERS

The classification of our equilibrium phases was done by examining orientational and positional ordering. To measure the long-range orientational order, we diagonalised the following second-rank symmetric tensor:

$$\mathbf{Q}^{\lambda\lambda} = \frac{1}{2N_p}\left\langle \sum_{i=1}^{N_p} \left(3\hat{\lambda}_i \cdot \hat{\lambda}_i - \mathbf{I}\right) \right\rangle \tag{1}$$

where $i$ indicates a generic particle, $\hat{\lambda} = \hat{x}, \hat{y}, \hat{z}$ denotes its unit orientation vector along its length ($L$), width ($W$) and thickness ($T$), respectively, $\mathbf{I}$ is the second-rank unit tensor, and the angular brackets denote ensemble average. When diagonalised, the tensor $\mathbf{Q}^{\lambda\lambda}$ produces three eigenvalues $S_{2,L}, S_{2,W}$, and $S_{2,T}$ and their corresponding eigevectors $\hat{\mathbf{n}}, \hat{\mathbf{m}}$, and $\hat{\mathbf{l}}$. For example, the tensor $\mathbf{Q}^{zz}$ is related to the largest eigenvalue $S_{2,L}$, and corresponding eigenvector $\hat{\mathbf{n}}$, which provides alignment along the particle axis $\hat{x}$. The calculation of the eigenvalues $S_{2,L}, S_{2,W}$ and $S_{2,T}$, referred to as uniaxial order parameters, allows to distinguish between an isotropic phase, where all eigenvalues vanish, and an ordered phase, where at least one of the eigenvalues is significantly larger than zero. We arbitrarily set the formation of a uniaxial LC phase when one of the three uniaxial order parameters is at least 0.40 (see Table I). To assess the system biaxiality, the biaxial order parameter for each axes can be calculated using the same symmetric tensor. To this end, the following equation is applied:

$$B_{2,L} = \frac{1}{3}\left(\hat{\mathbf{m}} \cdot \mathbf{Q}^{xx} \cdot \hat{\mathbf{m}} + \hat{\mathbf{l}} \cdot \mathbf{Q}^{yy} \cdot \hat{\mathbf{l}} - \hat{\mathbf{m}} \cdot \mathbf{Q}^{yy} \cdot \hat{\mathbf{m}} - \hat{\mathbf{l}} \cdot \mathbf{Q}^{xx} \cdot \hat{\mathbf{l}}\right) \tag{2}$$

The other two biaxial order parameters, $B_{2,W}$ and $B_{2,T}$, can be calculated from similar expressions. To determine $B_2$, it is sufficient to monitor the fluctuations of axes perpendicular to the main nematic director. For instance, if $S_{2,L}$ is the dominant uniaxial order parameter, it is sufficient to monitor $B_{2,L}$ as it indicates the fluctuations along axes $\hat{y}$ and $\hat{z}$ in the planes of $\hat{\mathbf{m}}$ and $\hat{\mathbf{l}}$. Table I shows the criteria to determine the symmetry of the phases observed in this work and consistent with Ref [1]. In Tables II to VIII, we report the uniaxial and biaxial order parameters of HBPs with $\sigma_L = 0.00$ to $\sigma_L = 0.30$.

TABLE I: Criteria of uniaxial and biaxial order parameters used in the classification of HBPs.

| $S_{2,L}$ | $S_{2,T}$ | $S_{2,W}$ | $B_{2,L}$ or $B_{2,T}$ | Phase |
|---|---|---|---|---|
| 0.00 - 0.20 | 0.00 - 0.20 | 0.00 - 0.20 | - | Isotropic |
| 0.40 - 1.00 | 0.00 - 0.35 | 0.00 - 0.35 | 0.00 - 0.30 | Uniaxial prolate |
| 0.00 - 0.35 | 0.40 - 1.00 | 0.00 - 0.35 | 0.00 - 0.30 | Uniaxial oblate |
| 0.40 - 1.00 | 0.35 - 1.00 | 0.35 - 1.00 | 0.20 - 0.35 | Weak biaxial prolate |
| 0.35 - 1.00 | 0.40 - 1.00 | 0.35 - 1.00 | 0.20 - 0.35 | Weak biaxial oblate |
| 0.40 - 1.00 | 0.40 - 1.00 | 0.40 - 1.00 | 0.35 - 1.00 | Biaxial |

TABLE II: Reduced pressure, packing fraction, uniaxial and biaxial order parameters for HBPs of $\sigma_L = 0.05$. Absolute errors are less than $5 \times 10^{-3}$.

| Phase | $P^*$ | $\eta$ | $S_{2,T}$ | $S_{2,W}$ | $S_{2,L}$ | $B_2$ |
|---|---|---|---|---|---|---|
| I | 0.060 | 0.265 | 0.039 | 0.019 | 0.032 | 0.012 |
| I | 0.070 | 0.286 | 0.057 | 0.021 | 0.061 | 0.021 |
| I | 0.075 | 0.297 | 0.086 | 0.023 | 0.073 | 0.025 |
| $N_U^-$ | 0.080 | 0.310 | 0.402 | 0.070 | 0.195 | 0.026 |
| $N_U^+$ | 0.085 | 0.326 | 0.257 | 0.131 | 0.577 | 0.037 |
| $N_U^+$ | 0.090 | 0.344 | 0.264 | 0.164 | 0.726 | 0.043 |
| $N_U^+$ | 0.095 | 0.375 | 0.265 | 0.247 | 0.894 | 0.042 |
| $Sm_U^+$ | 0.100 | 0.391 | 0.239 | 0.233 | 0.912 | 0.009 |
| $Sm_U^+$ | 0.110 | 0.415 | 0.261 | 0.256 | 0.926 | 0.035 |
| $Sm_U^+$ | 0.120 | 0.437 | 0.248 | 0.233 | 0.931 | 0.009 |
| $Sm_U^+$ | 0.130 | 0.458 | 0.261 | 0.233 | 0.921 | 0.011 |
| $Sm_U^+$ | 0.140 | 0.477 | 0.275 | 0.245 | 0.931 | 0.018 |
| $Sm_U^+$ | 0.160 | 0.513 | 0.312 | 0.243 | 0.907 | 0.068 |
| $Sm_U^+$ | 0.180 | 0.546 | 0.307 | 0.243 | 0.915 | 0.053 |
| $Sm_B^+$ | 0.200 | 0.573 | 0.431 | 0.331 | 0.888 | 0.208 |

TABLE III: Reduced pressure, packing fraction, uniaxial and biaxial order parameters for HBPs of $\sigma_L = 0.10$. Absolute errors are less than $5 \times 10^{-3}$.

| Phase | $P^*$ | $\eta$ | $S_{2,T}$ | $S_{2,W}$ | $S_{2,L}$ | $B_2$ |
|---|---|---|---|---|---|---|
| I | 0.060 | 0.265 | 0.052 | 0.019 | 0.052 | 0.024 |
| I | 0.070 | 0.288 | 0.081 | 0.024 | 0.068 | 0.023 |
| I | 0.075 | 0.297 | 0.114 | 0.024 | 0.090 | 0.027 |
| $N_U^-$ | 0.080 | 0.314 | 0.595 | 0.109 | 0.218 | 0.007 |
| $N_U^+$ | 0.090 | 0.342 | 0.231 | 0.160 | 0.752 | 0.004 |
| $N_U^+$ | 0.100 | 0.373 | 0.271 | 0.238 | 0.863 | 0.039 |
| $Sm_U^+$ | 0.105 | 0.395 | 0.260 | 0.249 | 0.921 | 0.022 |
| $Sm_U^+$ | 0.110 | 0.410 | 0.248 | 0.246 | 0.923 | 0.016 |
| $Sm_U^+$ | 0.120 | 0.429 | 0.249 | 0.249 | 0.940 | 0.018 |
| $Sm_U^+$ | 0.130 | 0.449 | 0.251 | 0.243 | 0.923 | 0.015 |
| $Sm_U^+$ | 0.140 | 0.478 | 0.265 | 0.243 | 0.940 | 0.019 |
| $Sm_U^+$ | 0.160 | 0.507 | 0.398 | 0.328 | 0.901 | 0.180 |
| $Sm_B^+$ | 0.180 | 0.539 | 0.455 | 0.385 | 0.913 | 0.223 |
| $Sm_B^+$ | 0.200 | 0.564 | 0.562 | 0.474 | 0.893 | 0.387 |

4TABLE IV: Reduced pressure, packing fraction, uniaxial and biaxial order parameters for HBPs of $\sigma_L = 0.15$. Absolute errors are less than $5 \times 10^{-3}$.

| Phase | $P^*$ | $\eta$ | $S_{2,T}$ | $S_{2,W}$ | $S_{2,L}$ | $B_2$ |
|---|---|---|---|---|---|---|
| I | 0.060 | 0.268 | 0.053 | 0.026 | 0.048 | 0.016 |
| I | 0.070 | 0.292 | 0.269 | 0.051 | 0.184 | 0.048 |
| $N_U^-$ | 0.080 | 0.319 | 0.602 | 0.148 | 0.282 | 0.080 |
| $N_U^-$ | 0.090 | 0.343 | 0.757 | 0.183 | 0.261 | 0.024 |
| $N_U^+$ | 0.100 | 0.371 | 0.290 | 0.231 | 0.830 | 0.044 |
| $N_U^+$ | 0.110 | 0.398 | 0.260 | 0.209 | 0.840 | 0.032 |
| $Sm_U^+$ | 0.120 | 0.421 | 0.357 | 0.283 | 0.848 | 0.111 |
| $Sm_U^+$ | 0.130 | 0.445 | 0.384 | 0.284 | 0.833 | 0.151 |
| $Sm_U^+$ | 0.140 | 0.465 | 0.405 | 0.293 | 0.824 | 0.183 |
| $Sm_U^+$ | 0.150 | 0.484 | 0.389 | 0.287 | 0.843 | 0.156 |
| $Sm_B^+$ | 0.160 | 0.503 | 0.557 | 0.434 | 0.843 | 0.360 |
| $Sm_B^+$ | 0.180 | 0.540 | 0.672 | 0.566 | 0.866 | 0.536 |

TABLE V: Reduced pressure, packing fraction, uniaxial and biaxial order parameters for HBPs of $\sigma_L = 0.18$. Absolute errors are less than $5 \times 10^{-3}$.

| Phase | $P^*$ | $\eta$ | $S_{2,T}$ | $S_{2,W}$ | $S_{2,L}$ | $B_2$ |
|---|---|---|---|---|---|---|
| I | 0.060 | 0.266 | 0.061 | 0.024 | 0.058 | 0.024 |
| I | 0.070 | 0.289 | 0.259 | 0.051 | 0.182 | 0.039 |
| $N_U^-$ | 0.080 | 0.314 | 0.642 | 0.131 | 0.219 | 0.018 |
| $N_U^+$ | 0.090 | 0.341 | 0.258 | 0.174 | 0.723 | 0.029 |
| $N_U^+$ | 0.100 | 0.362 | 0.258 | 0.205 | 0.812 | 0.032 |
| $N_U^+$ | 0.110 | 0.387 | 0.281 | 0.233 | 0.861 | 0.037 |
| $N_U^+$ | 0.120 | 0.411 | 0.413 | 0.316 | 0.840 | 0.113 |
| $N_B^+$ | 0.130 | 0.433 | 0.491 | 0.375 | 0.837 | 0.295 |
| $N_B$ | 0.140 | 0.455 | 0.641 | 0.498 | 0.822 | 0.465 |
| $N_B$ | 0.150 | 0.481 | 0.670 | 0.520 | 0.825 | 0.485 |
| $Sm_B^-$ | 0.160 | 0.496 | 0.910 | 0.588 | 0.667 | 0.521 |
| $Sm_B^-$ | 0.180 | 0.546 | 0.955 | 0.647 | 0.683 | 0.552 |
| $Sm_B^-$ | 0.200 | 0.573 | 0.948 | 0.646 | 0.692 | 0.567 |

TABLE VI: Reduced pressure, packing fraction, uniaxial and biaxial order parameters for HBPs of $\sigma_L = 0.20$. Absolute errors are less than $5 \times 10^{-3}$.

| Phase | $P^*$ | $\eta$ | $S_{2,T}$ | $S_{2,W}$ | $S_{2,L}$ | $B_2$ |
|---|---|---|---|---|---|---|
| I | 0.040 | 0.225 | 0.045 | 0.023 | 0.037 | 0.018 |
| I | 0.050 | 0.249 | 0.054 | 0.023 | 0.039 | 0.013 |
| I | 0.060 | 0.271 | 0.072 | 0.026 | 0.067 | 0.017 |
| I | 0.063 | 0.280 | 0.091 | 0.027 | 0.045 | 0.037 |
| I | 0.065 | 0.284 | 0.157 | 0.022 | 0.108 | 0.049 |
| $N_U^-$ | 0.070 | 0.297 | 0.402 | 0.067 | 0.186 | 0.027 |
| $N_U^+$ | 0.080 | 0.321 | 0.314 | 0.162 | 0.564 | 0.079 |
| $N_U^+$ | 0.090 | 0.348 | 0.286 | 0.196 | 0.714 | 0.047 |
| $N_U^+$ | 0.100 | 0.371 | 0.304 | 0.226 | 0.774 | 0.094 |
| $N_U^+$ | 0.105 | 0.384 | 0.364 | 0.269 | 0.772 | 0.131 |
| $N_B$ | 0.110 | 0.400 | 0.558 | 0.362 | 0.701 | 0.372 |
| $N_B$ | 0.120 | 0.419 | 0.584 | 0.376 | 0.720 | 0.395 |
| $N_B$ | 0.130 | 0.441 | 0.592 | 0.408 | 0.751 | 0.417 |
| $N_B$ | 0.140 | 0.465 | 0.643 | 0.517 | 0.813 | 0.459 |
| $N_B$ | 0.150 | 0.486 | 0.658 | 0.523 | 0.808 | 0.517 |
| $N_B$ | 0.160 | 0.505 | 0.709 | 0.569 | 0.816 | 0.579 |
| $N_B$ | 0.180 | 0.542 | 0.727 | 0.553 | 0.798 | 0.566 |
| $Sm_B^-$ | 0.200 | 0.575 | 0.792 | 0.608 | 0.796 | 0.668 |

TABLE VII: Reduced pressure, packing fraction, uniaxial and biaxial order parameters for HBPs of $\sigma_L = 0.25$. Absolute errors are less than $5 \times 10^{-3}$.

| Phase | $P^*$ | $\eta$ | $S_{2,T}$ | $S_{2,W}$ | $S_{2,L}$ | $B_2$ |
|---|---|---|---|---|---|---|
| I | 0.040 | 0.226 | 0.044 | 0.021 | 0.040 | 0.017 |
| I | 0.050 | 0.249 | 0.055 | 0.026 | 0.050 | 0.020 |
| I | 0.060 | 0.276 | 0.086 | 0.030 | 0.106 | 0.027 |
| I | 0.063 | 0.281 | 0.124 | 0.030 | 0.120 | 0.041 |
| I | 0.065 | 0.285 | 0.122 | 0.029 | 0.140 | 0.047 |
| $N_U^-$ | 0.070 | 0.300 | 0.457 | 0.109 | 0.201 | 0.030 |
| $N_U^-$ | 0.075 | 0.314 | 0.581 | 0.129 | 0.223 | 0.037 |
| $N_U^-$ | 0.080 | 0.324 | 0.592 | 0.143 | 0.269 | 0.072 |
| $N_U^-$ | 0.090 | 0.350 | 0.750 | 0.203 | 0.278 | 0.065 |
| $N_U^-$ | 0.095 | 0.361 | 0.787 | 0.200 | 0.260 | 0.042 |
| $N_B^+$ | 0.100 | 0.374 | 0.548 | 0.344 | 0.654 | 0.222 |
| $N_B^+$ | 0.110 | 0.398 | 0.616 | 0.386 | 0.663 | 0.301 |
| $N_B$ | 0.120 | 0.423 | 0.635 | 0.423 | 0.687 | 0.443 |
| $N_B$ | 0.130 | 0.446 | 0.690 | 0.540 | 0.780 | 0.457 |
| $N_B$ | 0.140 | 0.463 | 0.664 | 0.510 | 0.788 | 0.518 |
| $N_B$ | 0.160 | 0.507 | 0.671 | 0.508 | 0.783 | 0.524 |
| $N_B$ | 0.180 | 0.542 | 0.682 | 0.498 | 0.767 | 0.530 |
| $N_B$ | 0.200 | 0.570 | 0.718 | 0.541 | 0.783 | 0.574 |

8TABLE VIII: Reduced pressure, packing fraction, uniaxial and biaxial order parameters for HBPs of $\sigma_L = 0.30$. Absolute errors are less than $5 \times 10^{-3}$.

| Phase | $P^*$ | $\eta$ | $S_{2,T}$ | $S_{2,W}$ | $S_{2,L}$ | $B_2$ |
|---|---|---|---|---|---|---|
| I | 0.040 | 0.227 | 0.047 | 0.021 | 0.052 | 0.021 |
| I | 0.050 | 0.253 | 0.088 | 0.026 | 0.076 | 0.021 |
| I | 0.060 | 0.280 | 0.154 | 0.028 | 0.180 | 0.041 |
| $N_U^-$ | 0.068 | 0.300 | 0.466 | 0.010 | 0.202 | 0.031 |
| $N_U^-$ | 0.070 | 0.304 | 0.515 | 0.125 | 0.229 | 0.051 |
| $N_U^-$ | 0.080 | 0.328 | 0.629 | 0.115 | 0.263 | 0.055 |
| $N_U^-$ | 0.090 | 0.355 | 0.733 | 0.182 | 0.255 | 0.046 |
| $N_U^-$ | 0.095 | 0.366 | 0.727 | 0.209 | 0.306 | 0.054 |
| $N_B^+$ | 0.100 | 0.381 | 0.684 | 0.365 | 0.545 | 0.359 |
| $N_B$ | 0.110 | 0.402 | 0.633 | 0.418 | 0.674 | 0.467 |
| $N_B$ | 0.120 | 0.426 | 0.719 | 0.497 | 0.692 | 0.477 |
| $N_B$ | 0.130 | 0.450 | 0.704 | 0.542 | 0.768 | 0.561 |
| $N_B$ | 0.140 | 0.467 | 0.727 | 0.571 | 0.779 | 0.603 |
| $N_B$ | 0.160 | 0.509 | 0.748 | 0.587 | 0.785 | 0.609 |
| $N_B$ | 0.180 | 0.548 | 0.827 | 0.658 | 0.806 | 0.684 |

## II. PAIR CORRELATION FUNCTION

To distinguish nematic from smectic phases, we calculated the pair correlation function parallel to the nematic director(s), which formally reads:

$$g_{\parallel}(r_{\parallel}) = \frac{1}{N_p \rho S_{\parallel}} \left\langle \sum_i \sum_{j \neq i} \delta(r_{\parallel} - r_{\parallel,ij}) \right\rangle \tag{3}$$

where $\rho = N_p/V$, $N_p$ is the number of particles, $V$ the box volume, $\delta$ is the Dirac delta. $S_{\parallel}$ is a surface resulting from the intersection of a sphere with radius half of the simulation box and a plane perpendicular to the nematic director at a distance $r_{\parallel}$ from the center of the sphere. To practically evaluate it in our simulations, we employed the following expression [3]:

$$g_{\parallel}(r_{\parallel}) = \frac{1}{N_p \rho V_{\parallel}} \frac{N(r_{\parallel})}{N_c} \tag{4}$$

where $N_c$ is the number of sampled configurations. $N(r_{\parallel})$ is the number of times that the distance between two particles projected along the nematic director, $r_{\parallel,ij}$, is in the interval $|r_{\parallel} + \Delta r_{\parallel}|$. Finally, $V_{\parallel}$ is the volume of this region, that can be calculated as

$$V_{\parallel} = \pi \left( R^2 \Delta r_{\parallel} - 1/3((r_{\parallel} + \Delta r_{\parallel}/2)^3 - (r_{\parallel} - \Delta r_{\parallel}/2)^3) \right), \tag{5}$$

with $R$ the half of the shorter side of the simulation box. In Fig. I, we present the parallel pair correlation functions for the prolate nematic, oblate nematic, biaxial nematic, prolate uniaxial smectic, biaxial smectic of prolate symmetry and biaxial smectic of oblate symmetry.

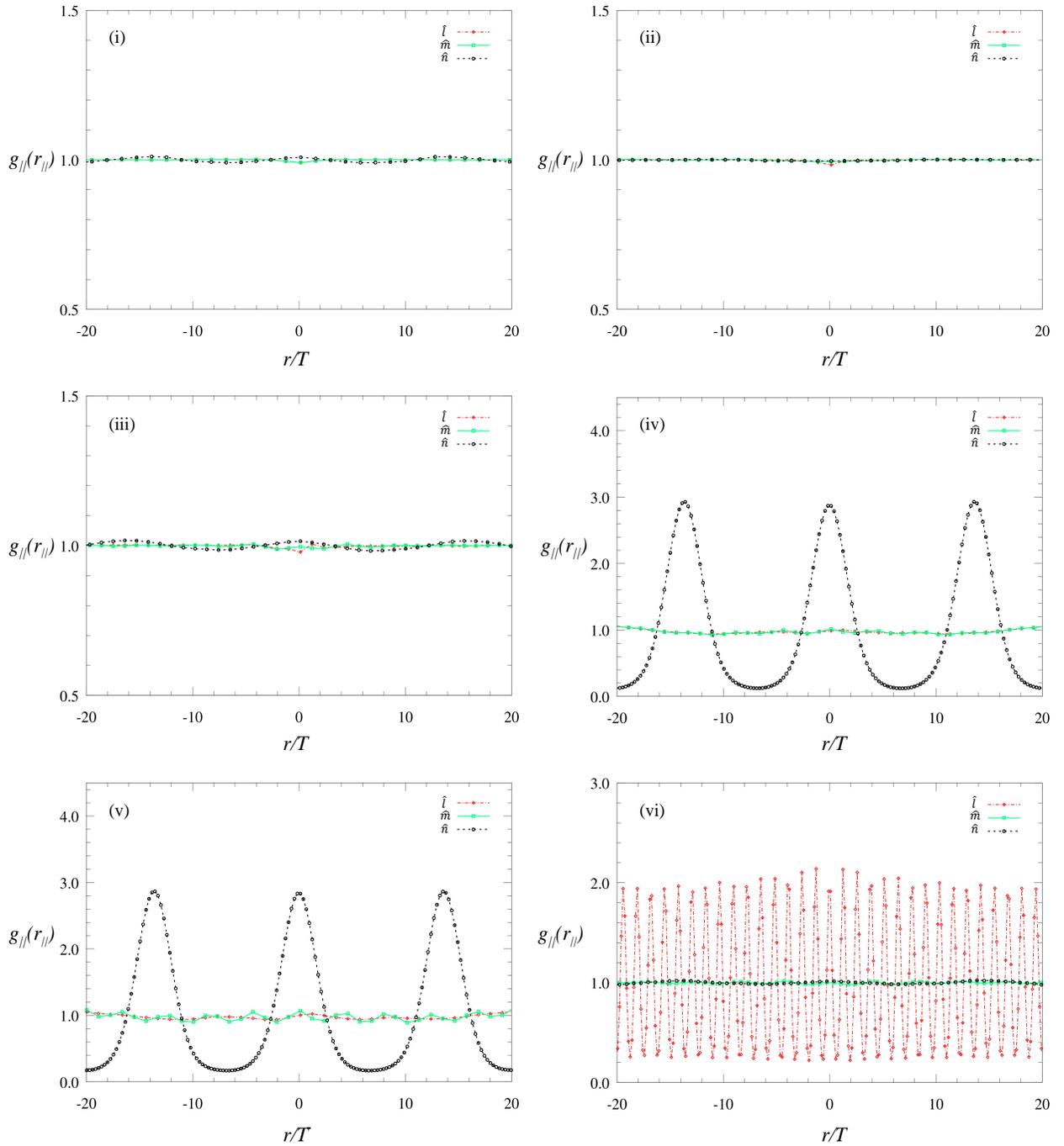

FIG. 1: Parallel pair distribution functions, $g_{\parallel}(r_{\parallel})$, along the nematic directors $\hat{\mathbf{l}}, \hat{\mathbf{m}}$ and $\hat{\mathbf{n}}$ of an ($i$) oblate nematic phase at $\sigma_L = 0.15$ and $\eta = 0.319$; ($ii$) prolate nematic phase at $\sigma_L = 0.10$ and $\eta = 0.342$; ($iii$) biaxial nematic phase at $\sigma_L = 0.25$ and $\eta = 0.487$; ($iv$) uniaxial smectic phase at $\sigma_L = 0.10$ and $\eta = 0.507$; ($v$) biaxial smectic phase with prolate symmetry at $\sigma_L = 0.10$; $\eta = 0.539$; and ($vi$) biaxial smectic phase with oblate symmetry at $\sigma_L = 0.18$ and $\eta = 0.546$.

## III. SMECTIC LAYER THICKNESS

Tables IX to XIII show the average thickness of smectic layers, defined by $\tau^* = \tau/T$ of HBPs with $\sigma_L = 0.05$ to $\sigma_L = 0.20$.

TABLE IX: Reduced pressure, packing fraction and average thickness of smectic layers for HBPs of $\sigma_L = 0.05$.

| $P^*$ | $\eta$ | $\tau^*$ |
|---|---|---|
| 0.100 | 0.391 | 14.1 |
| 0.110 | 0.415 | 14.0 |
| 0.120 | 0.437 | 13.7 |
| 0.130 | 0.458 | 13.5 |
| 0.140 | 0.477 | 13.5 |
| 0.160 | 0.513 | 13.3 |
| 0.180 | 0.546 | 13.3 |
| 0.200 | 0.573 | 13.2 |

TABLE X: Reduced pressure, packing fraction and average thickness of smectic layers for HBPs of $\sigma_L = 0.10$.

| $P^*$ | $\eta$ | $\tau^*$ |
|---|---|---|
| 0.105 | 0.395 | 14.4 |
| 0.110 | 0.410 | 14.3 |
| 0.120 | 0.429 | 14.2 |
| 0.130 | 0.449 | 14.1 |
| 0.140 | 0.478 | 13.9 |
| 0.160 | 0.507 | 13.7 |
| 0.180 | 0.539 | 13.7 |
| 0.200 | 0.564 | 13.6 |

TABLE XI: Reduced pressure, packing fraction and average thickness of smectic layers for HBPs of $\sigma_L = 0.15$.

| $P^*$ | $\eta$ | $\tau^*$ |
|---|---|---|
| 0.120 | 0.429 | 14.7 |
| 0.130 | 0.449 | 14.6 |
| 0.140 | 0.478 | 14.5 |
| 0.150 | 0.507 | 14.5 |
| 0.160 | 0.539 | 14.4 |
| 0.180 | 0.564 | 14.3 |

TABLE XII: Reduced pressure, packing fraction and average thickness of smectic layers for HBPs of $\sigma_L = 0.18$.

| $P^*$ | $\eta$ | $\tau^*$ |
|---|---|---|
| 0.160 | 0.496 | 1.3 |
| 0.180 | 0.546 | 1.3 |
| 0.200 | 0.573 | 1.3 |

TABLE XIII: Reduced pressure, packing fraction and average thickness of smectic layers for HBPs of $\sigma_L = 0.20$.

| $P^*$ | $\eta$ | $\tau^*$ |
|---|---|---|
| 0.20 | 0.575 | 1.3 |

## IV. FREE ENERGY CALCULATION

To gain an insight into the competition between prolate and oblate symmetries close to $I - N_U$ transition, we have compared the Helmholtz free energy of both $N_U^+$ and $N_U^-$ phases. To this end, we applied Onsager-like theory developed in our work for monodisperse systems, where the Helmholtz free energy is obtained as a combination of an entropy of mixing-like term and a correction for many-body interactions via virial expansions [2]. We stress that we are using a theory developed for monodisperse systems to calculate the free energy of polydisperse systems and hence our results should only be taken as a qualitative estimation of the free energy difference between the $N_U^+$ and $N_U^-$ phases at a particular value of $\eta$. Interested readers are referred to Ref [2] for additional details. The free energy calculation of our polydisperse systems are performed in the context of free particle rotation and assumes that the dominant uniaxial order parameter in the prolate ($S_{2,L}$) and oblate ($S_{2,T}$) nematic phases have the same value. The four virial coefficients are obtained from the $S_2$ calculated by Monte Carlo simulations (reported in Table II to Table VIII). The phases we consider in this analysis are the $N_U$ phases above the $I - N_U$ phase boundary for each polydispersity. For instance, at $\sigma_L = 0.05$, our simulation results indicate the formation of an $N_U$ phase at $\eta = 0.310$ with dominant uniaxial order parameter $S_2 = 0.402$. These values of $\eta$ and $S_2$ are used to obtain the corresponding virial coefficients and the entropy of mixing-like term for both $N_U^+$ and $N_U^-$ phases. In Fig. 2, the reduced free energy for $N_U^+$ and $N_U^-$ are shown for each size dispersity. For monodisperse systems, where $\sigma_L = 0.00$, the free energy of the $N_U^+$ phase is lower than the $N_U^-$ phase, suggesting that the system has a higher propensity to possess prolate symmetry, in agreement with Ref [2]. In polydisperse systems, the $N_U$ phases are generally more likely to form $N_U^+$ phases as well. However, the difference in free energy is such that relatively small density fluctuations, of the order of the thermal energy, might direct the systems to follow the oblate or prolate route, so determining their final symmetry. For $\sigma_L = 0.25$, we find that the system is more likely to form the $N_U^-$ phase, a tendency that is consistent with our simulation results.

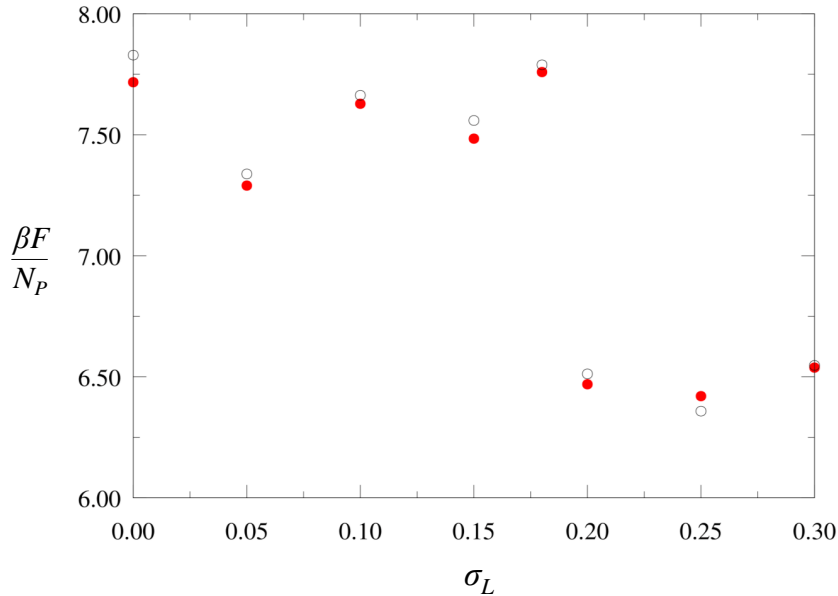

FIG. 2: Free energy per particle, plotted as a function of length polydispersity, $\sigma_L$. The parameter $\beta$ is the inverse temperature. Each point corresponds to the first $N_U$ phase that is observed above the I-N phase boundary. Solid and empty circles indicate, respectively, the free energy of prolate, $F_{N^+}^* \equiv \beta F_{N^+}/N_p$, and oblate, $F_{N^-}^* \equiv \beta F_{N^-}/N_p$ systems as obtained from theory.

With the same theory, the free energies $F_{N^+}$ and $F_{N^-}$ of $N_U$ phases deep in the $N_U$ region are also calculated. Tables XIV and XV show the estimated values of these free energies in monodisperse and polydisperse systems, respectively. Similar to what was observed for $N_U$ phases just above the I-N phase boundary, the free energy of prolate and oblate phases deep in the $N_U$ region are also very close to one another, suggesting an easy tendency for the systems to undergo director inversions with density fluctuations.

TABLE XIV: Reduced free energies of prolate, $F_{N+}^*$, and oblate, $F_{N-}^*$, uniaxial nematic phases deep in the $N_U$ region of monodisperse systems. Free energies highlighted in bold indicate the most stable phase.

| $P^*$ | $F_{N+}^*$ | $F_{N-}^*$ |
|---|---|---|
| 0.083 | **7.717** | 7.829 |
| 0.084 | 8.163 | **8.149** |
| 0.085 | **8.221** | 8.334 |
| 0.090 | 9.143 | **8.996** |

TABLE XV: Reduced free energies of prolate, $F_{N+}^*$, and oblate, $F_{N-}^*$, uniaxial nematic phases deep in the $N_U$ region of polydisperse systems. Free energies highlighted in bold indicate the most stable phase.

| | $\sigma_L$ | | | | | | | | | | | | | |
|---|---|---|---|---|---|---|---|---|---|---|---|---|---|---|
| $P^*$ | 0.05 | | 0.10 | | 0.15 | | 0.18 | | 0.20 | | 0.25 | | 0.30 | |
| | $F_{N+}^*$ | $F_{N-}^*$ | $F_{N+}^*$ | $F_{N-}^*$ | $F_{N+}^*$ | $F_{N-}^*$ | $F_{N+}^*$ | $F_{N-}^*$ | $F_{N+}^*$ | $F_{N-}^*$ | $F_{N+}^*$ | $F_{N-}^*$ | $F_{N+}^*$ | $F_{N-}^*$ |
| 0.068 | - | - | - | - | - | - | - | - | - | - | - | - | **6.597** | 6.606 |
| 0.070 | - | - | - | - | - | - | - | - | **6.469** | 6.512 | 6.420 | **6.358** | 6.642 | **6.568** |
| 0.075 | - | - | - | - | - | - | - | - | - | - | 7.095 | **7.079** | - | - |
| 0.080 | **7.290** | 7.338 | **7.628** | 7.663 | **7.484** | 7.559 | **7.759** | 7.789 | 7.742 | **7.660** | **7.713** | 7.749 | **7.660** | 7.764 |
| 0.085 | 7.374 | **7.229** | - | - | - | - | - | - | - | - | - | - | - | - |
| 0.090 | **8.787** | 8.917 | 9.035 | **8.963** | 8.933 | **8.862** | **8.694** | 8.822 | **8.745** | 8.808 | **8.630** | 8.757 | 8.925 | **8.839** |
| 0.095 | 10.074 | **9.936** | - | - | - | - | - | - | - | - | **9.327** | 9.392 | **9.271** | 9.410 |
| 0.100 | - | - | 10.038 | **9.876** | 10.053 | **10.023** | 9.656 | **9.480** | 9.718 | **9.700** | - | - | - | - |
| 0.105 | - | - | - | - | - | - | - | - | 10.683 | **10.603** | - | - | - | - |
| 0.110 | - | - | - | - | **11.112** | 11.315 | **10.919** | 10.956 | - | - | - | - | - | - |
| 0.120 | - | - | - | - | - | - | **12.289** | 12.518 | - | - | - | - | - | - |

## V.  EQUATION OF STATE

In Fig. 3, we report the $\eta$ vs $P^*$ equation of state for systems of HBPs with polydispersity index $\sigma_L = (0.00, 0.30)$. The reduced pressure, $P^*$, is defined as $P^* \equiv \beta P T^3$.

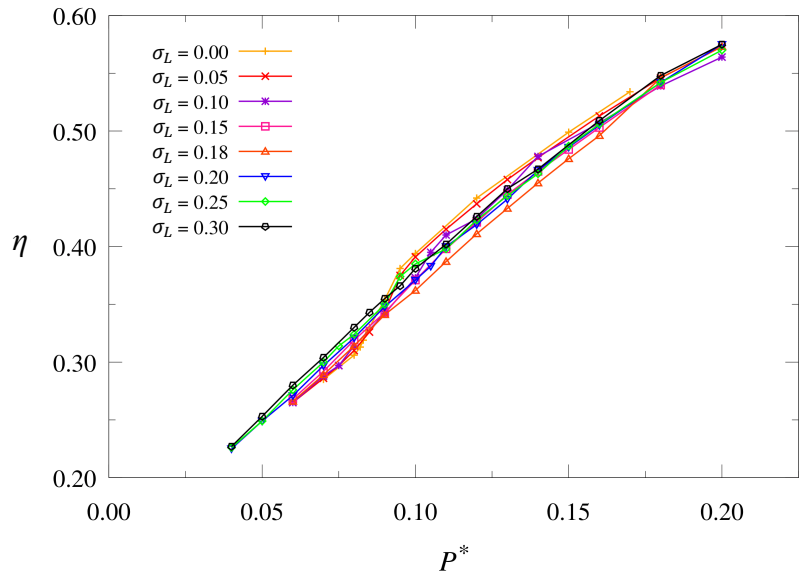

FIG. 3: Equation of state $\eta$ vs $P^*$ of HBPs at different values of $\sigma_L$. Solid lines are guides for the eyes.



\end{document}